\documentclass[aps,prb,twocolumn,showpacs,floatfix,superscriptaddress,amssymb,amsmath]{revtex4}
\usepackage{graphicx,times,epsfig,color}
\usepackage[all]{xy}
\usepackage{natbib}

\begin{document}

\def\salto{\vskip 1cm}
\def\lgr{\langle\langle}
\def\rgr{\rangle\rangle}

\title{Kondo temperature and screening extension in a double dot system.}

\author{L. C. Ribeiro}
\affiliation{Centro Federal de Educa\c{c}\~ao Tecnol\'ogica Celso Suckow da Fonseca (CEFET-RJ/UnED-NI), RJ, 26041-271, 
Brazil}

\author{E. Vernek}
\affiliation{Instituto de F\'isica, Universidade Federal de Uberl\^andia, Uberl\^andia 38400-902,
MG, Brazil}

\author{G. B. Martins}
\email[Corresponding author: ]{martins@oakland.edu}
\affiliation{Department of Physics, Oakland University, Rochester, Michigan 48309, USA}

\author{E. V. Anda}
\affiliation{Departamento de F\'{\i}sica, Pontif\'{\i}cia Universidade Cat\'olica do Rio de Janeiro, 22453-900, Brazil}

%


\begin{abstract}

In this work we use the Slave Boson Mean Field Approximation at finite U to study the effects of
spin-spin correlations in the transport properties of two quantum dots coupled in series 
to metallic leads. Different quantum regimes of this system
are studied in a wide range of parameter space. The main aspects related to the interplay between the half-filling 
Kondo effect and the antiferromagnetic correlation between the quantum dots are reviewed. Slave boson results for
conductance, local density of states in the quantum dots, and the renormalized energy
parameters, are presented. As a different
approach to the Kondo physics in a double dot system, the Kondo cloud extension inside the metallic
leads is calculated and its dependence with the inter-dot coupling is analyzed. In addition, the cloud extension
permits the calculation of the Kondo temperature  of the double quantum dot. This result is very similar to the corresponding critical
temperature $T_c$, as a function of the parameters of the system, as 
obtained by using the finite temperature extension of the Slave Boson Mean Field Approximation. 
\end{abstract}
\maketitle


\section{Introduction}

The Kondo regime in strongly correlated mesoscopic structures has been extensively studied since its
observation in a single quantum dot (QD) connected to metallic leads.\cite{Nature.391.156} The
increasing interest in investigating the underlying physics of these structures
is motivated by potential technological applications, such as the
design of single electron transistor devices based on the Coulomb blockade effect, as observed
in single QDs.\cite{nature01642,Applphys.97.031101}
A system composed of two tunnel coupled QDs connected in series with metallic leads 
[henceforth called a double QD (DQD)] is particularly interesting
because it is the simplest geometry in which the interplay of two energy scales determines its physical properties: 
({\it i}) the Kondo correlation between the spin of each QD and
the spins of the conduction electrons, and {\it (ii)} the antiferromagnetic correlation between the spins of
the two QDs. Each of these two regimes prevail in different regions of the parameter
space and compete in the crossover region. As discussed below, these regimes are manifested very clearly in charge transport
measurements.

Several experimental
and theoretical works have appeared in the last decade studying DQDs. The continued interest in DQDs 
stems from the early recognition \cite{PhysRevLett.61.125,PhysRevB.40.324} of a possible Non Fermi Liquid (NFL) Quantum Critical Point (QCP) 
separating a Fermi Liquid (FL) local singlet antiferromagnetic-phase from a Kondo screened FL Kondo-phase 
in the Two Impurity Kondo model (TIKM).\cite{note1}
Subsequent Numerical Renormalization group (NRG)\cite{RevModPhys.80.395} calculations on the Two Impurity Anderson model 
(TIAM) detailed the properties of this NFL QCP,\cite{JPSJ.61.2333} but already pointed out that the inter-impurity hopping 
suppresses the critical transition.\cite{note0} The great flexibility of mesoscopic systems, especially semiconducting 
lateral QDs, where a continuous tuning of the relevant 
physical parameters is possible, lead to a series of detailed theoretical studies of DQDs as a prototype for the TIAM and TIKM. 
Several papers used different flavors of 
the Slave Boson formalism to analyze either the TIAM \cite{JPSJ.67.1860,PhysRevB.63.125327,PhysRevLett.85.1946,PhysicaE.34.608,
PhysRevB.65.241304,PhysRevLett.89.136802} and the TIKM.\cite{PhysRevLett.82.3508,PhysRevLett.94.086602}
Detailed studies have also been done using NRG,\cite{JPSJ.61.2333,PhysRevB.62.10260} the Embedded Cluster Approximation (ECA),\cite{PhysRevB.62.9907} 
and the Non-Crossing Approximation.\cite{PhysRevB.67.245307} The results obtained confirmed that 
when the even-odd parity symmetry is broken, the critical transition is replaced by a crossover. In addition, in the TIAM, 
it was found that, as the inter-dot hopping increases, a coherent superposition of the many-body Kondo states of
each QD (forming bonding and antibonding combinations) results in a splitting of the Kondo resonance, which 
leads to a splitting of the zero bias anomaly in the differential conductance.\cite{PhysRevB.63.125327,PhysRevLett.85.1946,PhysRevB.62.9907} 
These many-body molecular states should not be confused with the single-particle molecular states (separated by 
an energy equal to twice the inter-dot hopping). In reality, the coherent single-particle molecular states 
were the first to be observed by R. H. Blick {\it et al.}\cite{PhysRevLett.80.4032} through 
careful analysis of the charging diagram of a DQD. Concurrently, by using a combination of charge transport and microwave 
spectroscopy, Oosterkamp {\it et al.}\cite{Nature.395.873} probed the formation of single-particle molecular states 
(which they called `covalent bonds') in a DQD by varying the tunneling coupling between the QDs, 
showing the possibility of controlling the quantum coherence in single-electron devices. Subsequently, 
Qin {\it et al.}\cite{PhysRevB.64.241302} showed that these coherent molecular states are robust even 
when coupled to acoustic phonons created in the system. The first observation of a coherent Kondo effect in a DQD was 
reported by Jeong {\it et al.}.\cite{Science.293.2221} Indeed, the splitting of the Kondo resonance (into bonding and 
anti-bonding many-body states) was then clearly observed, as had been theoretically 
predicted.\cite{PhysRevB.63.125327,PhysRevLett.85.1946,PhysRevB.62.9907} However, no {\it single} Kondo peak 
has been experimentally observed to date, probably due to the small values of inter-dot 
tunnel coupling required, which leads to a very small overall conductance at half-filling.

In addition, several recent papers have been trying to determine under what experimental conditions it would be possible to 
observe manifestations of the NFL QCP mentioned above. In the theoretical 
side, Affleck {\it et al.} \cite{PhysRevLett.102.047201} have used bosonization to determine the energy scale below which 
the RG flows to a FL QCP (away from the NFL QCP, as discussed above - 
see also Ref.~\onlinecite{PhysRevB.84.115111}). 
More importantly, an exact functional form for the conductance in the crossover region from NFL to FL was derived,\cite{PhysRevLett.102.047201} 
opening the doors for its possible experimental observation. Note that 
a very interesting experiment involving carefully approaching a Cobalt atom 
in an STM tip to another Cobalt atom laying on a gold surface has 
nonetheless failed to observe the NFL QCP physics.\cite{NPHYS2076} 
Also, M. Lee {\it et al.} have found through NRG a conduction-band mediated superexchange $J_I$, which competes with the 
direct superexchange term $J_U = 4t^2/U$ and dominates for large values of $U$ (intra-dot 
repulsion, and $t$ being the inter-dot hopping matrix element). These two terms 
(related to inter-lead charge transfer) are known to destabilize the NFL QCP, by breaking parity. 
Indeed, as mentioned above,\cite{note1} this QCP is closely associated to the TCK fixed point 
which is believed to have been observed in only one system.\cite{nature05556} 
In a recent paper, F. W. Jayatilaka {\it et al.},\cite{note1} 
using NRG, carefully analyze the experimental possibility of observing this QCP in a DQD.

In this work, we study the properties of a DQD (see Fig.~\ref{dp1}) using the Slave Boson Mean Field Approximation (SBMFA) 
at finite U. The main property we are interested in is the behavior of the
so-called {\it Kondo cloud} in a DQD. The Kondo cloud can
be understood as the spatial region occupied by the conduction electrons that are collectively involved in screening
the impurity spin.\cite{PhysRevB.53.9153,PhysRevB.77.180404} In the case of the DQD analyzed here (Fig.~\ref{dp1}),
at half-filling, when the tunnel coupling between
them is weak (as compared to the coupling of each QD to its adjacent lead), 
each QD will form its corresponding Kondo cloud with the conduction spins of the
lead to which it is directly connected. However, as the inter-dot coupling increases,
the system is driven through the crossover region between the Kondo regime and the
antiferromagnetic `molecular' regime. When passing through this region,
the Kondo cloud associated to each dot should `shrink' and disappear
accordingly. 

In a 1D system, the Kondo cloud is characterized by its length, which can be estimated  by
assuming that  the mean free path of the many-body quasi-particle is related to a time scale
associated to the Kondo temperature $T_{K}$. If one assumes that the electrons scattered by the single impurity
propagate with the Fermi velocity $v_{F}$, the Kondo cloud length, $\xi$, is estimated
as\cite{PhysRevB.53.9153}
\begin{eqnarray}\label{RK}
\label{rk}\xi&{\approx}&\frac{{\hbar}v_{F}}{k_{B}T_{K}},
\end{eqnarray}
where $\hbar$ and $k_B$ are Planck and Boltzmann constants, respectively.
Therefore, in the present case, it is important to determine whether for the DQD the
Kondo cloud length also scales with the inverse of $T_K$, as shown in Eq.~\ref{RK}.

From a theoretical point of view, the Kondo cloud has been studied through different
approaches.\cite{PhysRevB.81.045111} The most common one is based on the study of the dependence of the spin-spin
correlation with the distance between the impurity and the conduction electrons.\cite{PhysRevB.53.9153}
Using a variational approach it has been suggested that $\xi$ does not play a significant role in the
physics of a system of impurities in 2- and 3-dimensions.\cite{2007arXiv0708.3604S}
However, the Kondo cloud is an  important concept  to analyze the conductance 
properties of 1-dimensional systems, like the one studied in this paper.

Very recently, the authors studied the behavior of this correlation for arbitrary distances from a single 
impurity.\cite{PhysRevB.81.045111} This study has been done by analyzing the 
effect of the Kondo resonance on the local density of
states (LDOS) {\it away} from the impurity.
Using this approach, it was possible to show that $\xi$ behaves according to Eq.~\ref{RK}. As a consequence, it
is in principle possible to determine $T_K$ by studying the length of the associated Kondo cloud.

\begin{figure}
\label{dp1} \centering \vskip-1.0cm\hskip-0.0cm
\rotatebox{270}{\scalebox{0.33}{\includegraphics{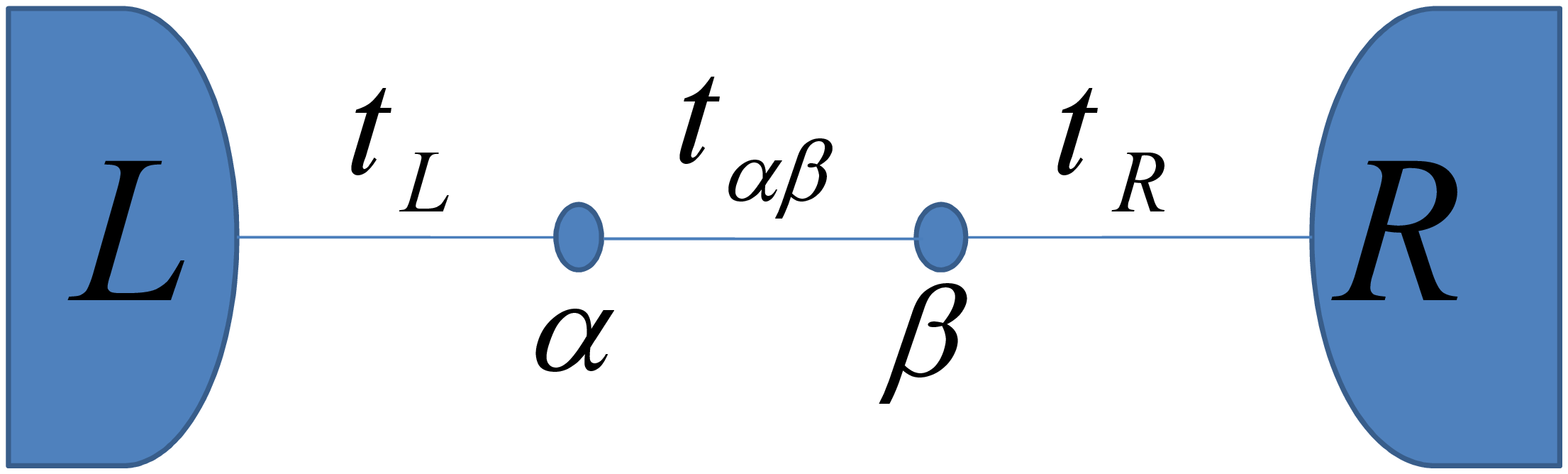}}}
\vskip-1.5cm \caption{(Color online) The figure sketches an artificial molecule consisting
of two QDs with intra-dot Coulomb interaction $U$, and tunnel coupled to each other
through the inter-dot matrix element $t_{\alpha\beta}$. In addition each QD is tunnel coupled 
to its adjacent metallic lead through the matrix element $t_{L(R)}$. 
The leads L and R act as charge reservoirs, being in thermodynamic equilibrium with the DQD.}
\label{dp1}
\end{figure}

To be more precise, the main objective of this paper is to answer the question: {\it how does the Kondo
temperature and the Kondo cloud  depend on the parameters of the DQD?} To this end,
we employ the finite-U SBMFA, which requires a more involved numerical
calculation than the infinite-U SBMFA,\cite{PhysRevLett.82.3508,PhysRevB.63.125327} 
and therefore eliminates the misconceptions
created by the artificial introduction of an extra antiferromagnetic inter-dot interaction $J$
in the Hamiltonian. We also study the system at finite
temperature and estimate $T_{K}$ by associating it to the temperature above which the width
of the Kondo peak in the LDOS vanishes. In this limit, within this approximation, the QDs
decouple from the rest of the system. We then compare the Kondo temperature obtained from the length
of the Kondo cloud with the one obtained from the slave-boson criterion just described
above. The results agree very well in the region of parameter space where the
system is in the Kondo regime. 

The paper is organized as follows: In Sec. II we present the model for the DQD and describe the finite-U SBMFA,
while in Sec. III we present SBMFA results at half-filling. Sec. IV is dedicated to the study of the Kondo cloud inside
the leads, presenting a way of calculating its extension and from it the Kondo temperature. In
Sec.~V we compare the results obtained for $T_K$ from the calculation of the Kondo cloud
extension with those obtained by calculating the width of the Kondo peak at finite temperature.
In Sec VI we present our conclusions, and relegate to the Appendix some more technical results.

\section{Model and Slave boson mean field approach}

The system of two quantum dots presented in the Fig.~\ref{dp1} is described by an Anderson
Hamiltonian which can be separated in three parts, $H=H_{0}+H_{t}+H_{leads}$,
where
\begin{eqnarray}
H_{0}&=&\sum_{i={\alpha},{\beta}\atop{\sigma}}\epsilon_ic^\dag_{i\sigma}c_{i\sigma}+\sum_{i=
{
\alpha},{\beta}}
Uc^\dag_{i\uparrow}c^\dag_{i\downarrow}c_{i\downarrow}c_{i\uparrow}
\end{eqnarray}
describes the isolated QDs, in which $c^\dag_{i\sigma}$ ($c_{i\sigma}$) creates (annihilates) an
electron with energy ${\epsilon}_{i}$ and spin $\sigma$ in the $i$th QD ($i=\alpha,~\beta$), and the second term
corresponds to the Coulomb interaction $U$ in each QD.
\begin{eqnarray}
H_{t}=\sum_{\sigma}\left(t_Lc^{\dag}_{-1\sigma}c_{\alpha\sigma}+t_Rc^{\dag}_{1\sigma}c_{\beta\sigma}
+H.c.\right)\nonumber\\
+\sum_{\sigma}t_{\alpha\beta}\left(c^{\dag}_{\alpha\sigma}c_{\beta\sigma}+H.c.\right)
\end{eqnarray}
describes the connection of each QD with its adjacent lead (first term) and the tunnel coupling 
between the QDs (second term). Finally,
\begin{eqnarray}
H_{leads}=t\sum^{\infty}_{i=1\atop{\sigma}}\left(c^{\dag}_{i\sigma}c_{i+1\sigma}+c^{\dag}_{-i\sigma}
c_ { -i-1\sigma}+H.c.\right)
\end{eqnarray}
describes the leads, modeled as two semi-infinity tight-binding chains of non-interacting sites
connected through the hopping term $t$. For simplicity,
we take the applied gate potential to be equal in both QDs, i.e.,
$\epsilon_\alpha=\epsilon_\beta=V_g$, and we consider symmetric coupling to the leads, 
$t_{L}=t_{R}=t^\prime$. Hereafter, we choose $t=1$ as
our energy unit and set $\hbar=k_B=1$.

Within the slave boson formalism, the physics underlying the Kondo regime is brought into the
model by the introduction of the auxiliary bosons $e_{i}$, $p_{i\sigma}$, and $d_{i\sigma}$,
which project the Hilbert space onto space sectors with zero, one, and two electrons, respectively.
To accommodate the new bosonic
field with these operators,  the Hilbert space is naturally enlarged, and the single electron
operator $c_{i\sigma}$ {($c^{\dag}_{i\sigma}$)} is replaced by a quasi-electron operator $Zc_{i\sigma}$
($Z^\dagger c_{i\sigma}^\dagger$), where the $Z$-operator, in mean field approximation becomes just a real
factor\cite{PhysRevLett.57.1362}
\begin{eqnarray}
Z_{i\sigma}&=&[1-{\langle}d_{i}{\rangle}^{2}-{\langle}p_{i\sigma}{\rangle}^{2}]^{-1/2}({
\langle} e_{i}{\rangle}{\langle}p_{i\sigma}{\rangle}+
{\langle}p_{i\sigma}{\rangle}{\langle}d_{i}{\rangle})\times\nonumber\\
&&\hspace{2.0cm}\times[1-{\langle}e_{i}{\rangle}^{2}-{\langle}p_{i\bar{\sigma}}{\rangle}^{2}]^{-1/2}
,
\end{eqnarray}
where $i$ denotes the $i$th QD and ${\langle}e_{i}{\rangle}$, ${\langle}p_{i\sigma}{\rangle}$, and
${\langle}d_{i\sigma}{\rangle}$ the mean value of the slave boson operators.
The full Hilbert space has to be restricted to the physically meaningful sector by imposing
the constraints
\begin{eqnarray}
{\langle}e_{i}{\rangle}^{2}+\sum_\sigma
{\langle}p_{i\sigma}{\rangle}^{2}+{\langle}d_{i}{\rangle}^{2}-1=0,
\end{eqnarray}
and
\begin{eqnarray}
n_{i\sigma}-{\langle}p_{i\sigma}{ \rangle}^{2}-{\langle}d_{i}{\rangle}^{2}=0,
\end{eqnarray}
via Lagrange multipliers $\lambda_{i}^{(1)}$ and   $\lambda_{i\sigma}^{(2)}$. Considering the
hybridization of the fermion operators in $H_{0}$ and $H_{t}$, and
also introducing the Lagrange multipliers ${\lambda}^{i}_{1}$ and ${\lambda}^{i}_{2\sigma}$, we can
write, the effective Hamiltonian as,

\begin{eqnarray}
H_{eff}&=&\sum_{{i={\alpha},{\beta}}\atop{\sigma}}{\tilde\epsilon}_{i}n_{i\sigma}+
\sum^{\sigma}_{i={L(j=\alpha)},\atop{{R(j=\beta)}}}t_{i}Z\left[c^{\dag}_{i\sigma}c_{j\sigma
} +H.c.\right]+
\nonumber\\
&+&\sum_{i={\alpha},{\beta}}U_{i}{\langle}d_{i}{\rangle}^{2}+\sum_{\sigma}t_{
\alpha\beta}Z^{2}\left[c^{\dag}_{\alpha\sigma}c_{\beta\sigma}+H.c.\right]+\nonumber\\
&+&{\sum_{i={\alpha},{\beta}}}{\lambda}^{i}_{1}\left[{\langle}e_{i}{\rangle}^{2}+
\sum_\sigma{\langle}p_{i\sigma}{\rangle}^{2}+{\langle}d_{i}{\rangle}^{2}-1\right]+\nonumber\\
&-&{\sum_{i={\alpha},{\beta}\atop{\sigma}}}{\lambda}^{i}_{2\sigma}\left[{\langle}p_{
i\sigma } {
\rangle}^{2}+{\langle}d_{i}{\rangle}^{2}\right]
+H_{leads},\hspace{0.5cm}
\end{eqnarray}
where $\tilde{\epsilon}_{i}={\epsilon}_{i}+{\lambda}^{i}_{2\sigma}$ is the renormalized
quasi-fermion energy.
It is important to notice that the replacement of the
single-fermion operator, in mean-field approximation,  results to be equivalent to
a renormalization of the connections $t^{\prime}$ and $t_{\alpha\beta}$ by a multiplicative parameter
$Z$ and $Z^{2}$, respectively.
With the effective Hamiltonian $H_{eff}$ we can obtain the free energy of the
system, which is minimized  with respect
to each component of the set of parameters,
\begin{eqnarray}
\label{888}
{\gamma}&=&\{e_{\alpha},e_{\beta},p_{\alpha\sigma},p_{\beta\sigma},d_{\alpha},d_{\beta},
{\lambda}^{\alpha}_{1},{\lambda}^{\beta}_{1},{\lambda}^{\alpha}_{2\sigma},
{\lambda}^{\beta}_{2\sigma}\}.
\label{888}
\end{eqnarray}
The minimization of the free energy provides a set of non-linear equations that has to be solved
in a self-consistent way. See Appendix B for details.

\section{Different Many-body regimes}

In this section, we present SBMFA numerical results for the low-temperature
physics of the DQD shown in Fig.~1. These results were obtained in the two different regions of
the parameter space, as described in the Introduction, and allow the analysis of transport properties in the two
well characterized quantum states of the DQD, as well as in the crossover region between them. 
In addition, some results for the molecular Kondo regime, which is characterized by an effective Coulomb
interaction $U_{eff}$, will be presented as an Appendix.

\subsection{Atomic Kondo Regime at Half-Filling}
To start the discussion concerning the different regimes of the system and the
crossover between them, we present in Figs.~\ref{dp2}, \ref{dp3}, and \ref{dp4} the renormalized
energy level $\tilde{\epsilon}_{\alpha(\beta)}$ and the local density 
of states (LDOS) of the QDs, the conductance $G$, and
the renormalization parameter $Z^{2}$, respectively. All these quantities (except for the LDOS in Fig.~\ref{dp2}(B) 
are calculated as a function of the gate potential $V_{g}$ and for different values of $t_{\alpha\beta}$,
and $E_{F}=0$ in both reservoirs. If the gate potential satisfies
$V_{g} \approx -U/2$ (i.e., the DQD is occupied by two electrons), then,
for small values of $t_{\alpha\beta}$, the system is expected to be in the Kondo regime resulting
from the singlet state created by the antiferromagnetic correlation of each QD with the
conduction electron spins of the corresponding lead. The formation of this Kondo state is
characterized by the plateau structure, around $V_g = -U/2$, at the Fermi level, observed in the results 
in Fig.~\ref{dp2}(A) for $t_{\alpha\beta}=0.025$, for example. This
plateau is related to the opening of a conducting channel (at $\omega=0$), which allow charge
transport through the DQD. Associated with this conduction channel there is a Kondo resonance at
the Fermi level, which can be clearly seen in Fig.~\ref{dp2}(B) [(black) square curve, for $t_{\alpha\beta}=0.025$],
which shows results for the QDs' local density of states (LDOS) close to the Fermi level for $V_{g}=-U/2$. Note
that the Kondo peak is split (and suppressed) for the smallest value of $t_{\alpha\beta}=0.025$ in Fig.~\ref{dp2}(B),
and that this splitting becomes more pronounced as $t_{\alpha\beta}$ increases.\cite{note2} Consequently,
the conductance (shown in Fig.~\ref{dp3}) is also suppressed from its maximum value $G = G_0$ at $V_{g}=-U/2$.
As mentioned in section II, another important characteristic of the Kondo regime, in the finite-U slave boson
approach, is the renormalization of the couplings through the parameter $Z$ (between both QDs
and between each QD and its adjacent lead), as shown in Fig.~\ref{dp4}. Note that for $t_{\alpha\beta}=0.025$
[(black) square curve], this renormalization is the strongest. This result shows
a large reduction of the parameter $Z$, reaching $Z{\approx}\sqrt{0.05}{\approx}0.22$, strongly suppressing the
renormalized hoppings $t^\prime$ and $t_{\alpha\beta}$, by factors $Z$ and $Z^2$, respectively [see eq.~(8)]. This is 
an indication of the formation of a Kondo state, where spin fluctuations are enhanced and charge fluctuations 
in the QDs are suppressed, which is reflect in the decrease of the effective hoppings connecting the QDs to each 
other and to the leads. One should remark that the split Kondo peak [(black) square curve] in Fig.~\ref{dp2}(B) 
is associated to the many-body coherent states which had been theoretically 
predicted\cite{PhysRevB.63.125327,PhysRevLett.85.1946,PhysRevB.62.9907} and experimentally observed.\cite{Science.293.2221} 

\begin{figure}
\label{dp2} \centering  \vskip-1.0cm\hskip-0.0cm
\rotatebox{0}{\scalebox{0.45}{\includegraphics{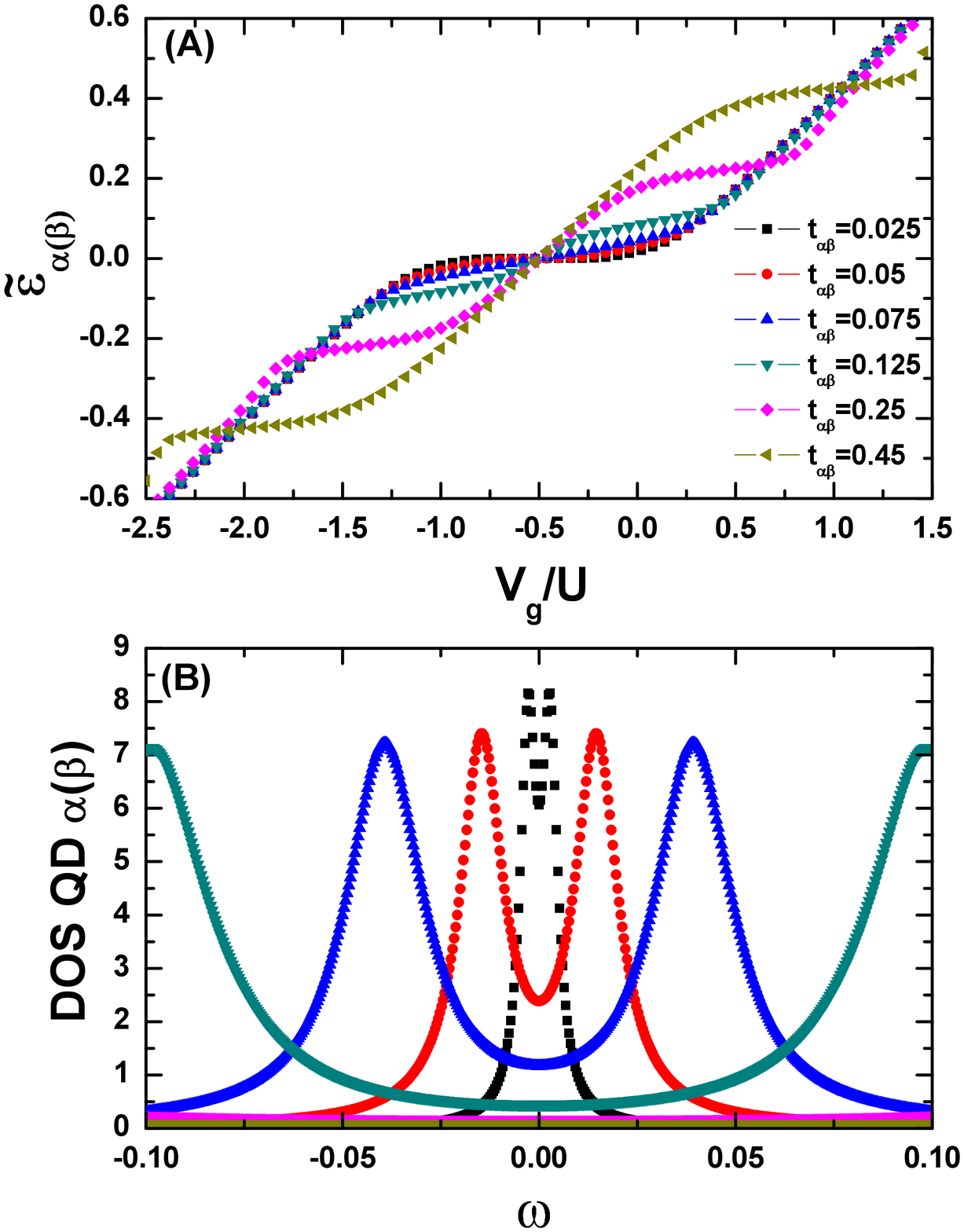}}}
\vskip-1.0cm \caption{(Color online) (A) behavior of the renormalized
local energy state $\tilde{{\epsilon}}_{\alpha(\beta)}$ of the QDs as a function of the gate potential
$V_{g}$. The parameters used are $U=0.5$, $t'=0.15$, Fermi energy $E_{f}=0$, and with different magnitudes
for $t_{\alpha\beta}$ (see legend). As $t_{\alpha\beta}$ increases, 
the plateau observed at $\tilde{{\epsilon}}_{\alpha(\beta)}=0$, characteristic of the Kondo regime, 
is gradually suppressed. Note that for the largest value of $t_{\alpha\beta}=0.45$ 
one has now two plateaus located at $\tilde{{\epsilon}}_{\alpha(\beta)} \approx \pm t_{\alpha\beta}$, which 
(as noted in Appendix A) are associated to the molecular Kondo regimes at quarter-filling. 
(B) QD's LDOS for $V_{g}=-U/2$, and the same $t_{\alpha\beta}$ values as in panel (A). 
For the smallest value of $t_{\alpha\beta}=0.025$ [(black) squares curve) a splitting of the Kondo resonance
is already observed. Note that for $t_{\alpha\beta}=0.125$ [(green) down-triangles curve], the separation between the peaks is already close 
to $2t_{\alpha\beta}$, which is the separation expected for single-particle molecular orbitals, as mentioned 
in the Introduction.}
\label{dp2}
\end{figure}
\subsection{Molecular Regime at Half-Filling}
With the gradual increase in the magnitude of the connection $t_{\alpha\beta}$,
an antiferromagnetic interaction $J={t_{\alpha\beta}}^2/U$ develops between the spins localized in each QD,
competing with the Kondo state and, for large
values of $t_{\alpha\beta}$, it is responsible for the total suppression of the Kondo state at half-filling. 
From Fig.~\ref{dp2}(A) we note the progressive destruction, as $t_{\alpha\beta}$
increases, of the plateau in the $\tilde{\epsilon}_{\alpha(\beta)} ~ vs ~ V_g$ curves, around half-filling. 
Simultaneously, it is possible to observe the development of two other plateaus, for $V_{g}/U>0$ and $V_{g}/U<-1$, that correspond to 
the emergence of one- and three-electron Kondo regimes, respectively, which will be discussed in Appendix A.
In Fig.~\ref{dp2}(B) (where $V_g=-U/2$, i.e., particle-hole symmetric point),
the transition to the molecular regime, accompanied by the destruction of the Kondo resonance
[(black) squares curve), and the formation of the molecular antiferromagnetic state
[(blue) triangle curve) is reflected in the LDOS ${\rho}_{\alpha(\beta)}$ of the QDs. 
Note in Fig.~\ref{dp2}(B) that for $t_{\alpha\beta}=0.125$ [(green) down-triangle curve], the separation 
between the peaks is already $\approx 2t_{\alpha\beta}$, indicating that these peaks 
are associated to the coherent single-particle molecular states, as experimentally observed 
in Refs.~\onlinecite{PhysRevLett.80.4032,Nature.395.873,PhysRevB.64.241302}.
The transition to the molecular regime can also be observed in Fig.~\ref{dp4},
showing that the renormalization of the dot hopping connections is reduced, as the
renormalization parameter rapidly approaches the value $Z{\approx}1.0$ as $t_{\alpha\beta}$
increases. This effect is reflected in the conductance of the system as presented in Fig.~\ref{dp3},
which is strongly suppressed at half-filling for larger values of $t_{\alpha\beta}$.
The plateaus outside the half-filling regime shown in Fig.2(A), for the largest values of $t_{\alpha \beta}$ 
correspond to the conductance of the DQD at the molecular one- and three-Kondo regime that, as mentioned, 
are discussed in Appendix A.

\begin{figure}
\label{dp3} \centering  \vskip-1.0cm\hskip-0.0cm
\rotatebox{270}{\scalebox{0.33}{\includegraphics{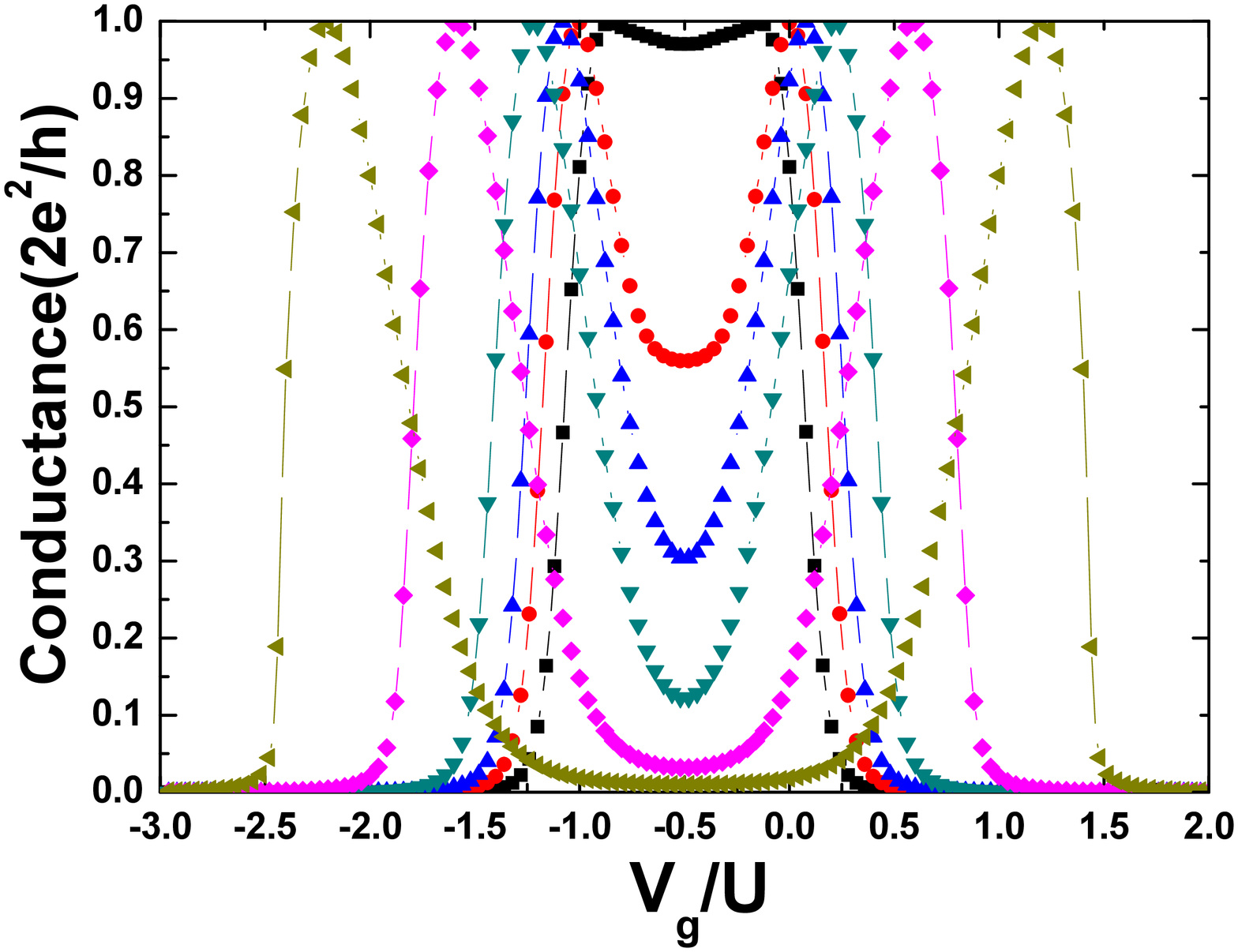}}}
\vskip0.0cm \caption{(Color online) Conductance as a function of $V_g$ for $t_{\alpha\beta}=0.025$
[(black) squares curve], $t_{\alpha\beta}=0.05$ [(red) circles], $t_{\alpha\beta}=0.075$ [(blue) up-triangles],
$t_{\alpha\beta}=0.125$ [(green) down-triangles], $t_{\alpha\beta}=0.25$ [(magenta) diamonds]
and $t_{\alpha\beta}=0.45$ [(light green) left-triangle]. The other parameters are $t^\prime= 0.15$
and $U = 0.5$.} \label{dp3}
\end{figure}
\begin{figure}
\label{dp4} \centering  \vskip-1.0cm\hskip-0.0cm
\rotatebox{270}{\scalebox{0.33}{\includegraphics{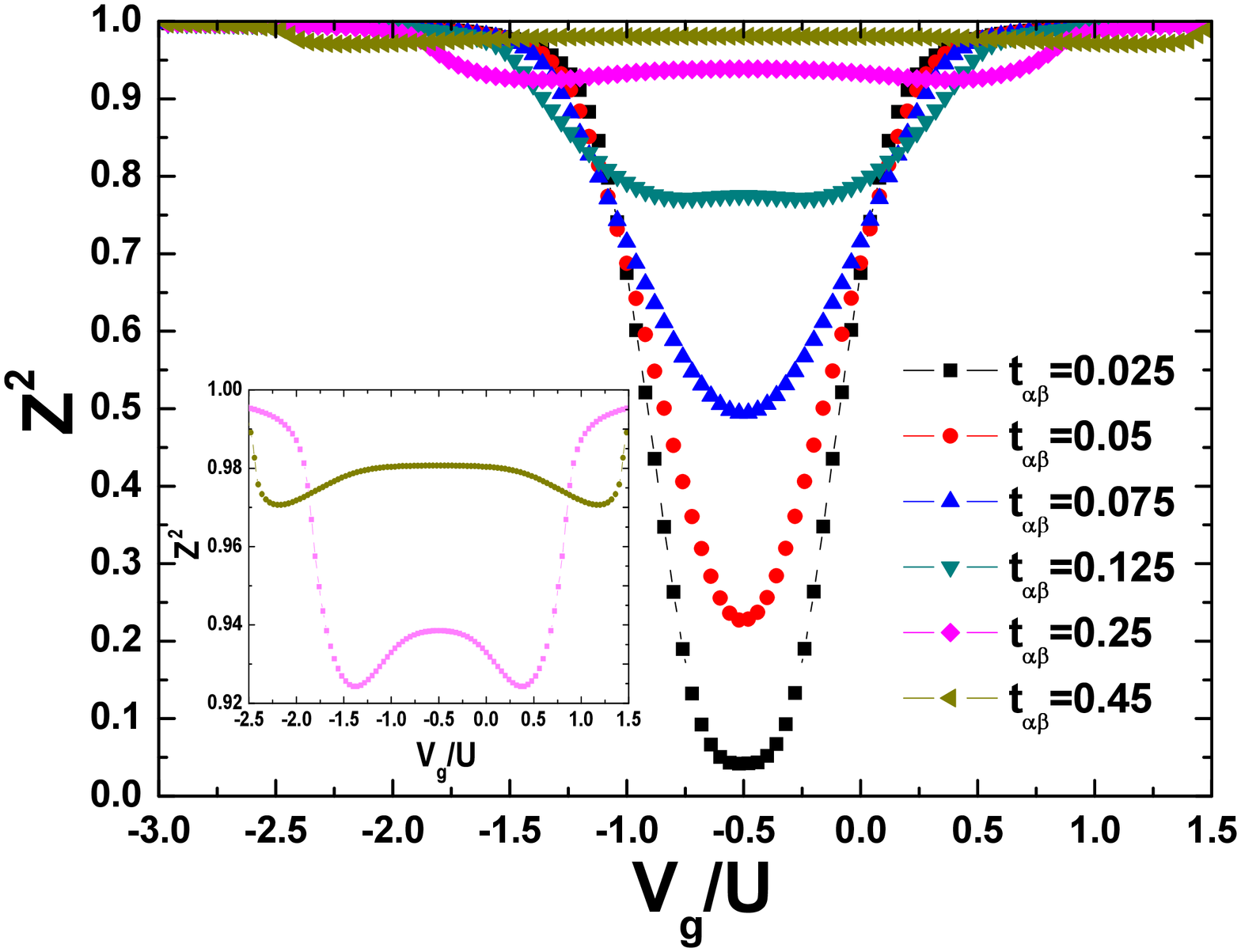}}}
\vskip0.0cm \caption{(Color online) Renormalization parameter $Z^{2}$ as a function of
$V_{g}$ for the same $t_{\alpha\beta}$, $U$, and $t^\prime$ values as in Fig.~\ref{dp3}.
The inset shows the small renormalization of $Z$ characteristic of the molecular Kondo regime.} \label{dp4}
\end{figure}

\section{Kondo Cloud}

\subsection{Cloud Extension Function}
In this section we introduce a new perspective to the Kondo problem in a strongly
coupled DQD at half-filling, based on the analysis of the extension of the Kondo cloud inside the metallic leads.
To analyze the extension of the Kondo cloud we use the method developed in Ref.~\onlinecite{PhysRevB.81.045111},
where the authors analyze the propagation into the leads (away from the QD) of the perturbation in the LDOS, introduced by the Kondo
resonance {\it at} the QD. To study this propagation, a function $F(N)$ was defined, which quantifies the
perturbation produced by the presence of the Kondo cloud in the $N$th site of the lead (counted from
the border of the semi-infinity chain). In the current work we consider a similar expression,
\begin{eqnarray}
\label{eqq11}
F(N)=\int_{-\infty}^{+\infty}[{\rho}_{N}^{k}(\omega)-
{\rho}_{N}^{nk}(\omega)]{\rho}_{\alpha(\beta)}(\omega)d\omega,
\end{eqnarray}
to quantify this perturbation. Instead of using a Gaussian function with width $T_K$ to convolute
the  density of states and eliminate the Friedel oscillations,\cite{PhysRevB.77.180404} allowing
the study of the region near the Fermi level, we use the LDOS ${\rho}_{\alpha(\beta)}$ of the QD.
Note that, as ${\rho}_{\alpha}={\rho}_{\beta}$, we will omit the subscripts $\alpha$ and $\beta$ from the rest
of the equations in this subsection.
The use of this definition for the convolution function ($\rho$) has the advantage of directly incorporating
into $F(N)$ the physical information associated with the Kondo ground state. In this expression, ${\rho}_{N}^{k}$ and
${\rho}_{N}^{nk}$ represent the LDOS calculated in the $N$th site inside the lead (L or R) with the
system {\it in} and {\it away from} the Kondo regime, respectively. The last condition is enforced by
disconnecting the DQD from the leads, i.e., by calculating the LDOS for $t^{\prime}=0$. It is well known that, within the SBMFA,
this is equivalent to taking the system to a temperature $T >T_{K}$.\cite{JPSJ.67.1860,PhysRevLett.82.3508}

The LDOS appearing in the integrand of $F(N)$ is proportional to the imaginary part of the Green's
function  $G_{NN}$ defined at the $N^{th}$ site inside the semi-infinite metallic leads. Considering
the left lead, for instance, we can write
\begin{eqnarray}
G_{NN}&=&{g}_{NN}+g_{N1}t^{\prime}G_{{\alpha}N}.
\end{eqnarray}
In this expression ${g}_{NN}$ is defined as the Green's function at site N of the left lead when
$t^{\prime}=0$ and $g_{N1}$ satisfies the equation
\begin{eqnarray}
g_{N1}&=&t^{N-1}\tilde{g}_{L}^{N},
\end{eqnarray}
where
\begin{eqnarray}
\tilde{g}_{L}=\frac{{\omega}-\sqrt{{\omega}^{2}-4t^{2}}}{2t^{2}}
\end{eqnarray}
is the Green's function defined for the left lead. For $G_{\alpha N(N\alpha)}$ we have
\begin{eqnarray}
G_{{\alpha}N}=G_{{N}\alpha}&=&g_{N1}t^{\prime}G_{\alpha\alpha},
\end{eqnarray}
where $G_{\alpha\alpha}$ is the dressed function at the QD $\alpha$.
Substituting $G_{{\alpha}N}$ into eq.~(11), yields
\begin{eqnarray}
G_{NN}&=&{g}_{NN}+t^{\prime2}(g_{N1})^{2}G_{\alpha\alpha}.
\end{eqnarray}
We note that the Kondo physics is introduced into $G_{NN}$ through the term proportional to
$t^{\prime2}$.  The imaginary part of this function is proportional to the LDOS ${\rho}^{k}_{N}$ at the
$N^{th}$ site. The non-Kondo solution is obtained by eliminating the effects resulting
from the presence of the QD by considering $t^{\prime}=0$
in the expression of $G_{NN}$. So, ${\rho}^{nk}_{N}$ is defined as
${\rho}^{nk}_{N}=-(1/\pi){\tt Im}\,{g}_{NN}$.

Considering the analytical expression for ${\rho}^{k}_{N}$ and ${\rho}^{nk}_{N}$ we obtain
\begin{eqnarray}
\label{eqq1}
{\rho}^{k}_{N}-{\rho}^{nk}_{N}=-\frac{1}{\pi}{\tt Im}[t^{\prime2}(g_{N1})^{2}G_{\alpha\alpha}].
\end{eqnarray}
Substituting $g_{N1}$ into eq.~(\ref{eqq1}) we obtain
\begin{eqnarray}
{\rho}^{k}_{N}-{\rho}^{nk}_{N}=-\frac{1}{\pi}{\tt Im}[t^{\prime2}
\tilde{g}_{L}^{2N}t^{2N-2}G_{\alpha\alpha}].
\end{eqnarray}
This expression, when substituted into eq.~(\ref{eqq11}), results in
\begin{eqnarray}
F(N)=-\frac{1}{\pi}\int_{-\infty}^{+\infty}{\tt
Im}[t^{\prime2}\tilde{g}^{2N}_{L}t^{2N-2}G_{\alpha\alpha}(\omega)]
{\rho}(\omega)d{\omega},
\end{eqnarray}
where all the effects of strong correlation present at site $N$ are contained in the Green's function of the QDs.
Note that when $t_{\alpha\beta}=0$ one should obtain the results from previous calculations for a
single-QD Kondo cloud.\cite{PhysRevB.81.045111} For finite $t_{\alpha\beta}$, the competition between
Kondo and antiferromagnetism is contained in $G_{\alpha\alpha}(\omega)$, and therefore should be reflected in $F(N)$.

\subsection{LDOS}

The LDOS ${\rho}_{N}^{k}$ and ${\rho}_{N}^{nk}$, calculated at the site $N=50$ of the leads, for the 
system with $t_{\alpha\beta}=0$, are shown in Fig.~\ref{dp10}. 
As in the rest of this paper, the results presented in this figure were
obtained for $V_{g}=-U/2$, $U=0.5$, and $t^{\prime}=0.2$. The
LDOS ${\rho}_{N}^{nk}$ was calculated for an isolated lead ($t^{\prime}=0$). The Kondo resonance at
the QD is shown in the (blue) triangle curve of Fig.~\ref{dp10}. On the other hand, the LDOS
${\rho}_{N=50}^{k}$ [(red) circle curve] shows a small peak at the Fermi level. The presence of this
peak is a sign of the Kondo resonance `propagating' through the sites of the leads and reflects
the existence of the Kondo cloud, with an extension ${\xi}$, dependent on the
Kondo temperature $T_{K}$. In our context, the extension of the Kondo cloud is obtained
from $F(N)$ (see next section for more details). It is important to note here that the propagated Kondo peak along the
leads appears as a resonance or an antiresonance depending on 
whether the site $N$ is even or odd, respectively.\cite{PhysRevB.81.045111} 

\begin{figure}
\label{dp10} \centering  \vskip-1.0cm\hskip-0.0cm
\rotatebox{270}{\scalebox{0.33}{\includegraphics{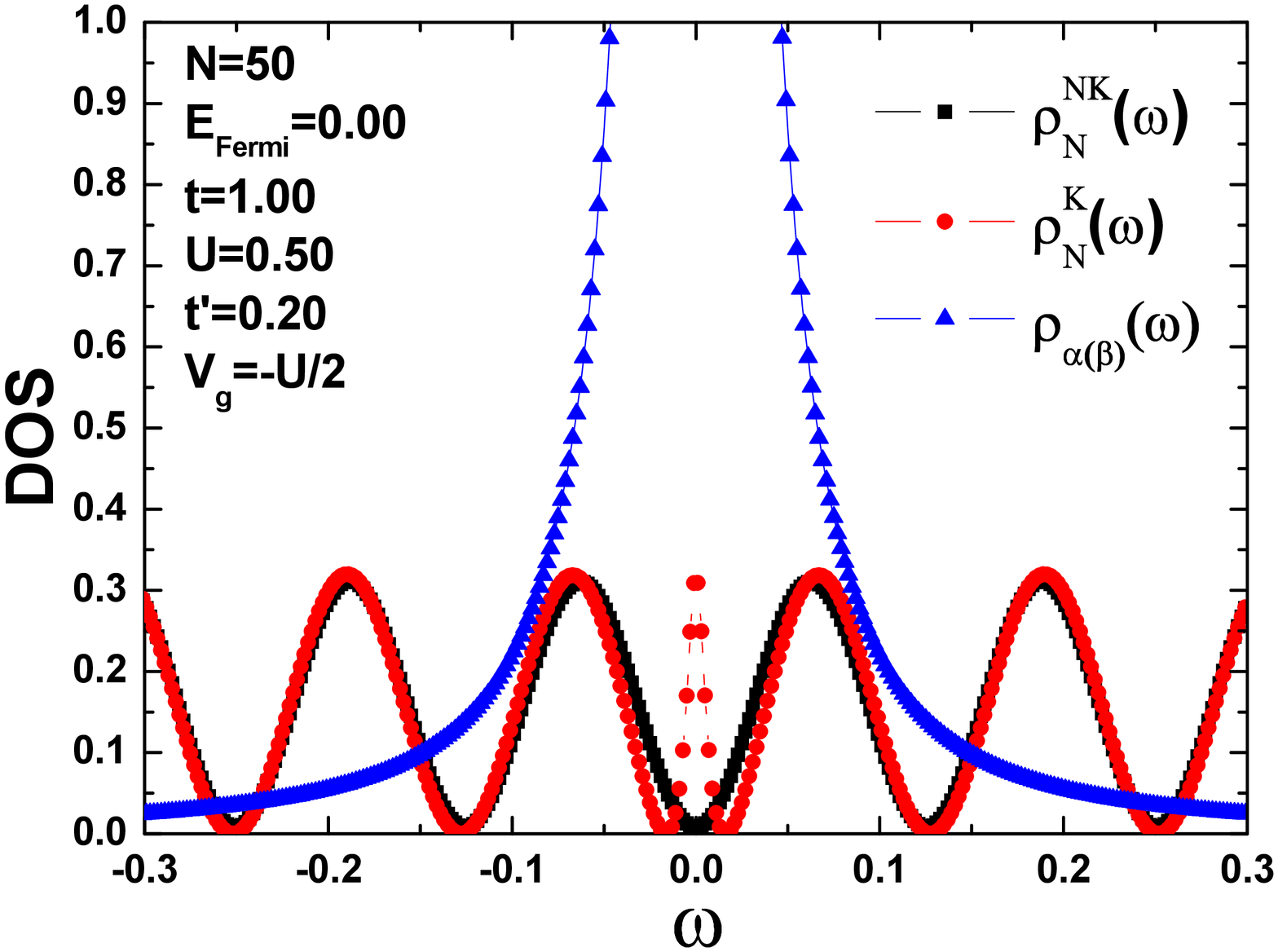}}}
\vskip0.0cm \caption{{(Color online) The figure shows the effect produced by the presence of the
impurity $\alpha(\beta)$ in the LDOS calculated at the site $N=50$ [inside the lead L(R)]. The (black) squares curve 
shows the LDOS (at site $N=50$) for the isolated lead; the (red) circles curve shows the LDOS
when each lead is connected to its respective QD, for $t_{\alpha\beta}=0$; 
and the (blue) triangles curve shows the LDOS for the impurity.}} \label{dp10}
\end{figure}

Figure \ref{dp11} shows the behavior of the LDOS for different
values of $t_{\alpha\beta}$, illustrating how the competition between the Kondo regime and the
antiferromagnetic correlation manifests itself in the LDOS at an arbitrary site ($N=50$).
The (magenta) dashed curve is the LDOS for a lead with no connection to the
dots, which, as a reference, corresponds  to the expected result obtained for a non-Kondo
regime. The (black) down-triangle curve corresponds to $t_{\alpha\beta}=0$ ($t^{\prime}=0.2$) and shows the
characteristic Kondo resonance propagated to $N=50$. It is clear from the figure that an increase
of the tunnel coupling $t_{\alpha\beta}$ drives the system from a Kondo regime to a non-Kondo ground state, 
as the LDOS at (and around) the Fermi level is clearly suppressed, approaching the value for the 
disconnected DQD [(magenta) dashed curve]. 

\section{Kondo Temperature}

\subsection{Kondo Temperature from the Cloud Extension Function}
\begin{figure}
\label{dp11} \centering  \vskip-1.0cm\hskip-0.0cm
\rotatebox{270}{\scalebox{0.33}{\includegraphics{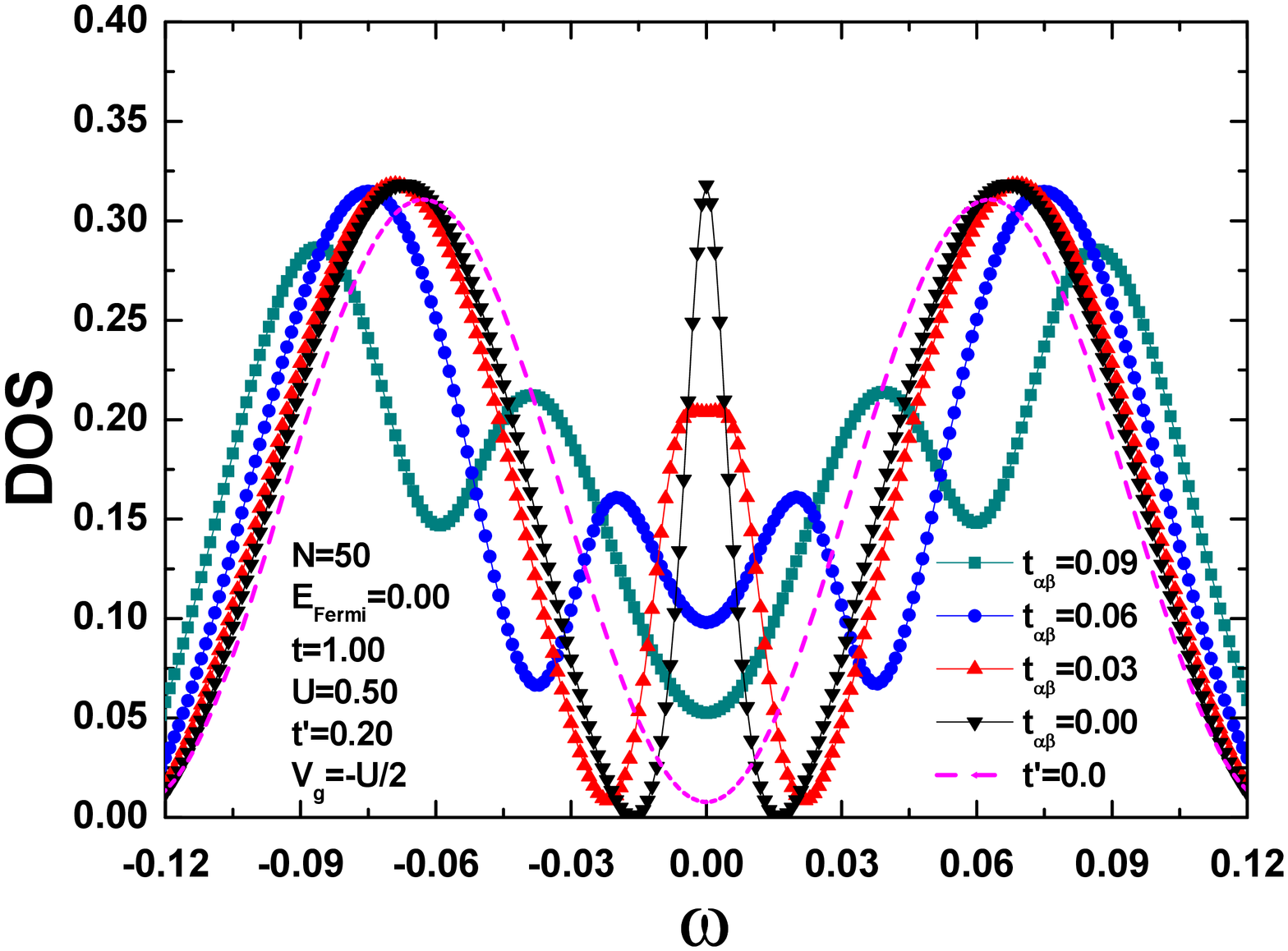}}}
\caption{{(Color online) The figure shows the effect produced by the antiferromagnetic
interaction $J=4t_{\alpha\beta}^{2}/U$ between the QDs in the LDOS at site $N=50$ inside the
leads. The (black) down-triangle curve shows the LDOS calculated at $N=50$ for $t_{\alpha\beta}=0$.
The (red) up-triangles, (blue) circles, and (cyan) down-triangles curves are obtained for
$t_{\alpha\beta}=0.03$, $t_{\alpha\beta}=0.06$, and $t_{\alpha\beta}=0.09$, respectively. The (magenta) dashed curve
shows the LDOS for the isolated leads ($t'=0$). The other parameters are $V_{g}=-U/2$, $U=0.5$, and $t'=0.2$.}} \label{dp11}
\end{figure}

Figure \ref{dp12} shows the logarithmic dependence of the function $F(N)$ for various inter-dot
couplings $t_{\alpha\beta}$. Differently from the result obtained for a system of one QD connected
to a metallic lead,\cite{PhysRevB.81.045111} this function presents now an oscillatory behavior with a
frequency associated to the inter-dot coupling. However, similarly to the single-QD case, the extension
${\xi}$ of the Kondo cloud can still be obtained, in this case by the exponential decay of its envelop
function. The physical information contained in this function is
extracted from the straight lines, tangent to the logarithm of the $F(N)$ function. Specifically,
the extension of the Kondo cloud, as a function of the inter-dot connection, can be obtained from the
slopes of these lines, which are proportional to $1/{\xi}$ (see eq.~(23) in Ref.~\onlinecite{PhysRevB.81.045111}).

\begin{figure}
\label{dp12} \centering  \vskip-1.0cm\hskip-0.0cm
\rotatebox{270}{\scalebox{0.33}{\includegraphics{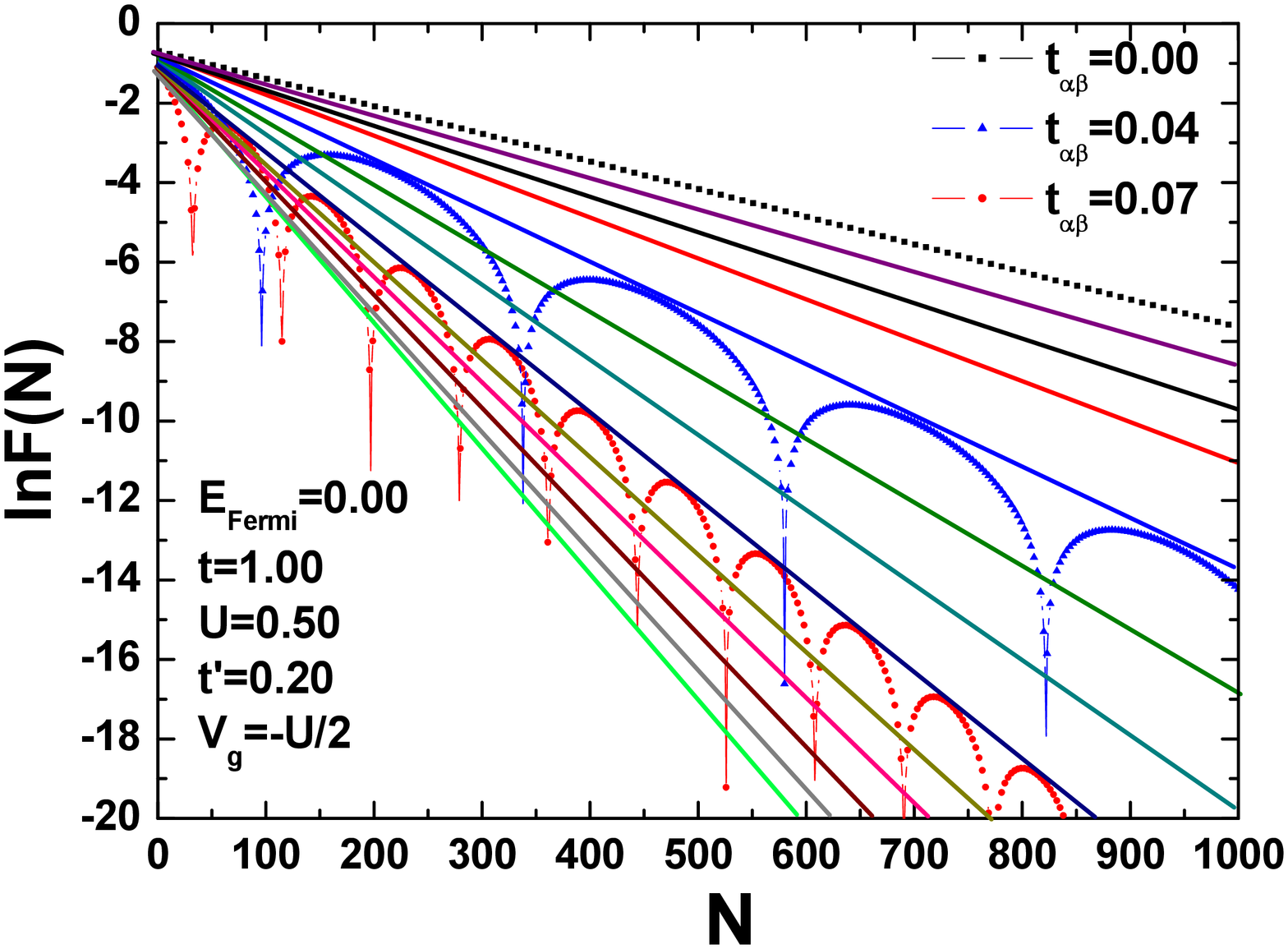}}}
\vskip0.0cm \caption{{(Color online) Natural logarithm of the cloud extension function $F$ {\it vs} 
the distance N (from the border of the semi-infinite chain) for 
$t_{\alpha\beta}=0.0$ [(black) squares], $t_{\alpha\beta}=0.04$ [(blue) triangles], and $t_{\alpha\beta}=0.07$ [(red) circles].
The straight lines are tangents to $lnF$ and correspond to $t_{\alpha\beta}=0.01$ [(purple) line], 
$t_{\alpha\beta}=0.02$ [(black) line], $t_{\alpha\beta}=0.03$ [(red) line], ..., $t_{\alpha\beta}=0.12$ 
[(green) line]. The other parameters are $V_{g}=-U/2$, $U=0.5$, and $t^{\prime}=0.2$.}} \label{dp12}
\end{figure}

In Fig.~\ref{dp12} we show the function $\ln F(N)$ for $t_{\alpha\beta}=0$ [(black) squares curve], 
$t_{\alpha\beta}=0.04$ [(blue) triangles curve], and $t_{\alpha\beta}=0.07$ [(red) circles curve], 
as well as their respective tangent lines. For intermediate values of $t_{\alpha\beta}$ 
we present only the tangents to $\ln F(N)$. As mentioned above, the slopes
of these straight lines are proportional to $1/{\xi}$ and allow us to obtain the Kondo temperature through the expression
$\xi=\frac{\Gamma}{T_{K}}$, proposed in Ref.~\onlinecite{PhysRevB.81.045111}. The values obtained for
the Kondo temperature $T_{K}$ are presented in the (red) circles curve in Fig.~\ref{dp14}, as function of
$t_{\alpha\beta}$ (we label it as $T_{K}^{KC}$, i.e., the Kondo temperature obtained 
through the extension of the Kondo cloud). We note that, for $\frac{4t^{2}_{\alpha \beta}}{U}<{T_K}$, which implies
$t_{\alpha\beta}<0.07$, $T_{K}$ presents an exponential behavior, in
accordance with results obtained by Aono and Eto\cite{PhysRevB.63.125327} using a slave boson
formalism with an infinite Hubbard $U$.

\subsection{Kondo Temperature in the Finite Temperature SBMFA}

In this section we determine the Kondo temperature $T_{K}$ of the DQD by extending the slave
boson  formalism for finite temperature. In this formalism, the impurity is decoupled from the rest
of the system for a critical temperature $T_c$. Although phase transitions are typical artifacts of mean
field solutions, the temperature for which this decoupling occurs can be considered as a reasonable
approximation to the Kondo temperature, which physically corresponds to a crossover between
two regimes. The decoupling occurs through the parameter $Z$, which renormalizes the
coupling of the QDs to the leads, $\tilde{t'}=Zt'$ and
$\tilde{t_{\alpha\beta}}=Z^{2}t_{\alpha\beta}$, which is vanishingly small when
$T \approx T_{K}$. The parameter $Z$,  as a function of temperature, is studied in 
Fig.~\ref{dp13}, for various values of $t_{\alpha\beta}$. 
We see from this result that the parameter $Z^{2}$ vanishes rapidly
when the temperature approaches the characteristic value $T_{c}$.

\begin{figure}
\label{dp13} \centering  \vskip-1.0cm\hskip-0.0cm
\rotatebox{270}{\scalebox{0.3}{\includegraphics{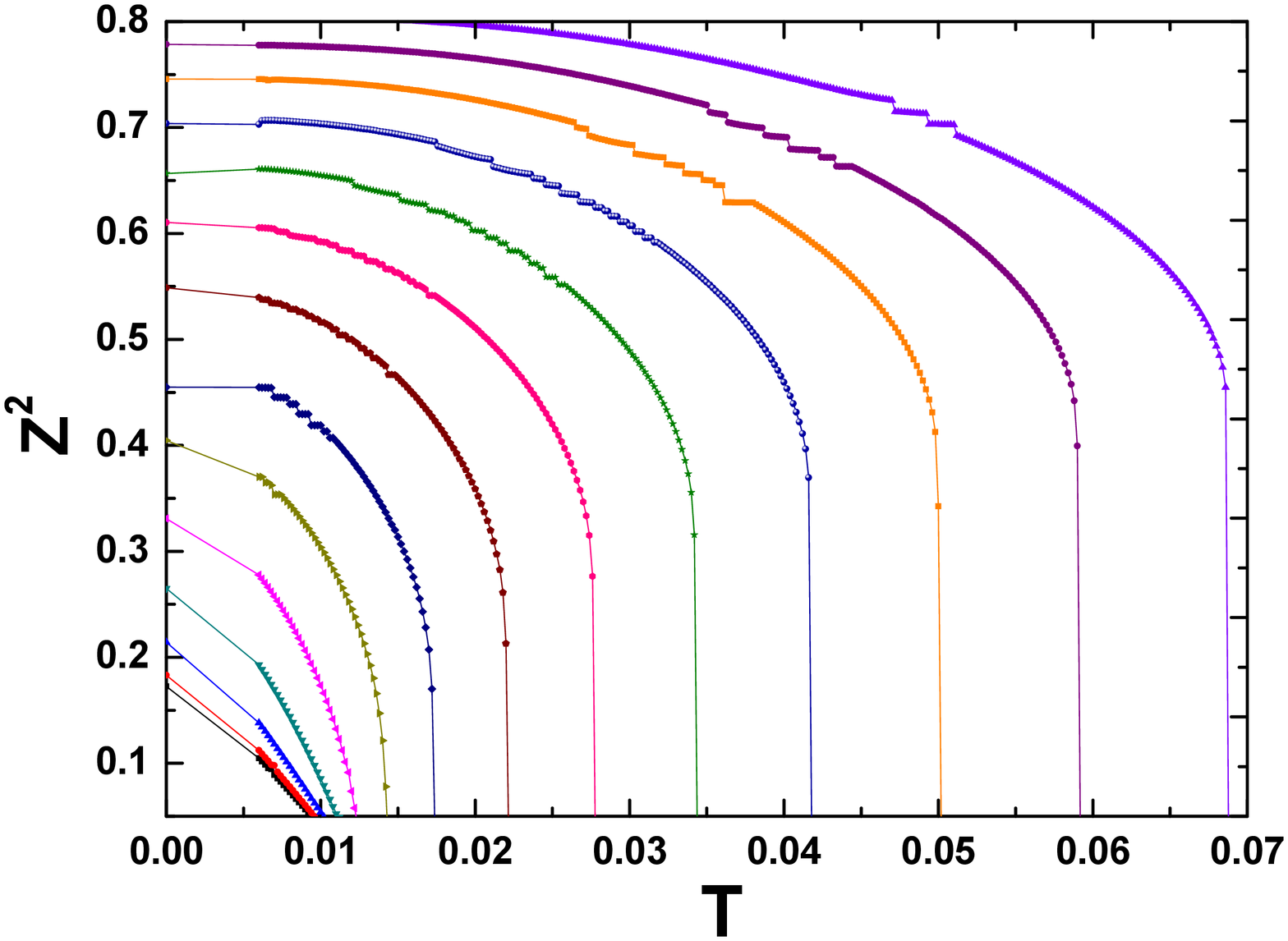}}}    
\vskip0.0cm \caption{(Color online) Renormalization parameter $Z^{2}$ as a function of 
 temperature $T$ for values of $t_{\alpha\beta}$ ranging from $0.00$ [(black) 
leftmost curve] to $0.13$ [(purple) rightmost curve]. The other parameters are 
$t^{\prime}=0.2$, $U=0.5$, $V_{g}=-U/2$, and $E_{f}=0.0$.} \label{dp13}
\end{figure}

The values of $T_{c}$ obtained for the different values of $t_{\alpha\beta}$ are represented by
the square (black) curve in the Fig.~\ref{dp14} (we label it $T_{K}^{Cut}$), and it agrees 
with the values obtained for $T_{K}^{KC}$ from the function $\ln F(N)$ associated to the extension of the Kondo cloud in the
metallic leads. We observe a qualitative and semiquantitative agreement between these 
two ways of calculating the Kondo temperature, giving support to the 
interpretations we are proposing. We note that  increasing $t_{\alpha\beta}>0.07$ the
values of $T_{c}$ become  increasingly different from the $T_K$ values obtained from the extension 
of the Kondo cloud. These are precisely values for which ${t_{\alpha\beta}}^2/U>T_K$, and the discrepancy starts at 
$t_{\alpha\beta}\gtrsim0.07$, where the DQD enters a crossover region which
extends up to values of $t_{\alpha\beta}$ for which ${4t_{\alpha\beta}^{2}}/{U}>2T_{K}^{(0)}$, 
where the DQD enters into an antiferromagnetic phase ($T_{K}^{(0)}$ is the Kondo temperature 
of each QD when $t_{\alpha\beta}=0.0$), according to previous works (see Ref.~\onlinecite{PhysRevB.63.125327} 
and references therein).

\section{CONCLUSIONS}

In this paper, the interplay between the antiferromagnetic interaction and the Kondo effect for a 
DQD at half-filling has been studied in detail. In particular, we have analyzed the dependence of the extension
of the Kondo cloud on the inter-dot tunnel coupling ($t_{\alpha\beta}$) when the system is driven from the half-filling 
Kondo regime to the antiferromagnetic molecular ground state, as $t_{\alpha\beta}$ increases. The extension
of the Kondo cloud and its associated $T_K$ were obtained by analyzing the propagation inside the leads
of the Kondo resonance `originating' at the dots. In the Kondo regime, the $T_K$ results obtained 
through the Kondo cloud extension were almost identical to the 
`Kondo transition temperature' $T_c$ obtained from the finite-temperature extension of the MFSBA. Although we know
that the decoupling of the DQD from the leads (occurring at $T_c$) is an artifact of the MFSBA, our results confirm that the
$T_c$ can be reliably taken to be $\approx T_K$ in a DQD, very much in the same way as in the case of one impurity.
Note that the dependence of $T_k$ obtained by both approaches agrees with previous results by Aono and Eto.\cite{PhysRevB.63.125327}

The study of the Kondo regime in the DQD was done within the context of a finite U
treatment. As U is finite, the Hamiltonian incorporates the correlation between the QD's spins. This
allows us to avoid introducing an artificial antiferromagnetic interaction between the QD's, normally
incorporated in $U=\infty$ approaches (as done, for example, in Ref.~\onlinecite{PhysRevLett.82.3508}), 
which provides more confidence to our results.
Its important to note that the study developed in this paper opens a set of conceptual ideas
that can be useful to understand the effects of the RKKY interaction between two QD's located at an
arbitrary distance between themselves, as well as the transport associated to them.

\begin{figure}
\label{dp14} \centering  \vskip-1.0cm\hskip-0.0cm
\rotatebox{270}{\scalebox{0.33}{\includegraphics{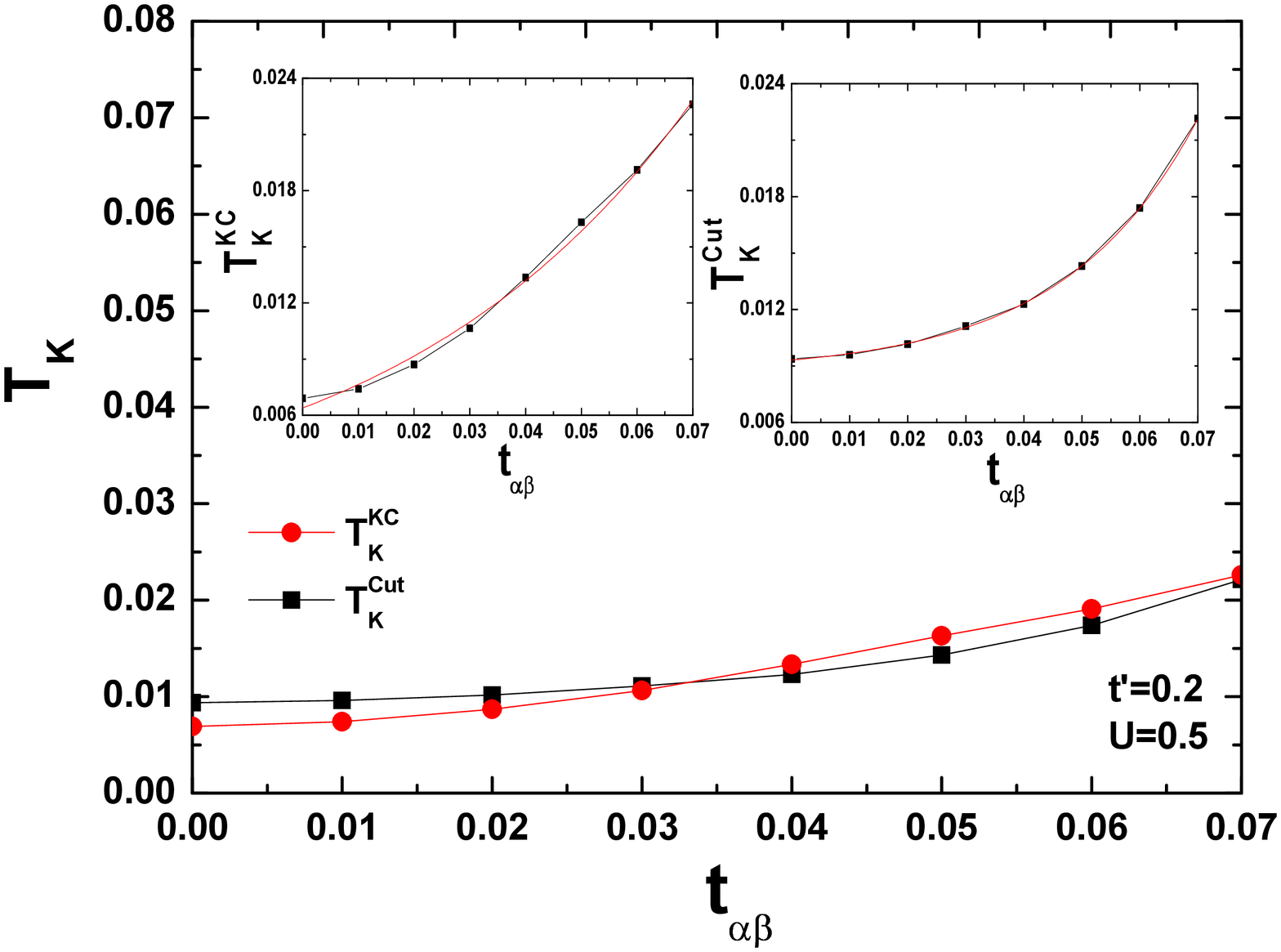}}}  
\vskip0.0cm \caption{{(Color online) Kondo temperature $T_{K}^{KC}$ and $T_{K}^{Cut}$
as a function of $t_{\alpha\beta}$. The (red) circles curve corresponds to $T_{K}$
obtained from the extension $\xi$ of the Kondo cloud, while the (black) squares 
curve corresponds to $T_{K}$ as obtained through the critical temperature
$T_{c}$ for which the QD's are decoupled (see text), with $Z\rightarrow0$. 
The insets show the exponential fits 
obtained for each $T_{K}$ in the region where the system is expected to be in the Kondo regime. 
The values for the other parameters are $V_{g}=-U/2$, $U=0.5$, and $t^{\prime}=0.2$.}}
\label{dp14}
\end{figure}

\appendix

\section{Molecular Kondo Regime at Quarter-Filling}

Given the importance in a DQD, as mentioned in the Introduction, of the concept of molecular
states (or molecular orbitals), the authors will briefly describe the
properties of the molecular orbitals in a DQD, as done previously
for related systems.\cite{PhysRevB.73.153307} For recent work
by some of the authors, see References \onlinecite{PhysRevB.78.085308,PhysRevB.80.035119}.
A system of two or more coupled QDs is said
to be in the molecular regime when the charge transport occurs through orbitals that
involve linear combinations of levels in each QD (which could be called `atomic' orbitals).
The molecular Kondo effect occurs when an unpaired spin residing in a molecular orbital is
screened by conduction electrons.\cite{PhysRevB.80.035119}
It is clear that, in the molecular Kondo regime, the charge at each
QD in a multi-QD system does not correspond to an integer value. As a consequence, the non-integer
QD charge ground state of a multi-QD structure cannot be used as a criterion to
identify the system as being in a fluctuating valence regime, as it could be in a molecular
Kondo ground state, involving more than one QD. The determination of the real nature of the ground state of these systems
requires a careful analysis, mainly for parameter values where the system
is in a transition between the molecular and the atomic regimes.

In the case here analyzed for a DQD, by adjusting $V_{g}$ to have one (or three) total number of 
electrons in the DQD, and considering relatively large
values of $t_{\alpha\beta}$, the quarter-filled molecular Kondo regime is accessed.
In the context of the finite-$U$ slave
boson formalism, this regime is characterized by two plateau structures in
$\tilde{\epsilon}_{\alpha(\beta)}$ [as observed at half-filling in Fig.~\ref{dp2}(A)] at
$\tilde{\epsilon}_{\alpha(\beta)}=\pm t_{\alpha\beta}$. The formation of these plateaus is
associated with two new resonances which allow charge transport
through the DQD, as can be seen by the well separated double peaks in the conductance  
in Fig.~\ref{dp3} [(yellow) left-triangle and (magenta) diamond curves, for $t_{\alpha\beta}=0.45$ 
and $0.25$, respectively]. 

Figure \ref{dp4} shows that as the system is driven out of the Kondo regime, at $V_g=-U/2$, by increasing
$t_{\alpha\beta}$, the  parameter $Z \rightarrow 1.0$, thus eliminating the renormalization of the hopping 
matrix elements [see eq.~(8)].  As an important characteristic of this result, we note two small suppressions
of $Z^2$ for the (yellow) left-triangle and (magenta) diamond curves in the one- and three-electron 
regions of $V_g$. This region is amplified in the inset of Fig.~\ref{dp4} and reflects a small renormalization
provided by the $Z$ parameter, characteristic of the molecular Kondo regime.
\begin{figure}
\label{condan} \centering  \vskip-1.0cm\hskip-0.0cm
\rotatebox{270}{\scalebox{0.33}{\includegraphics{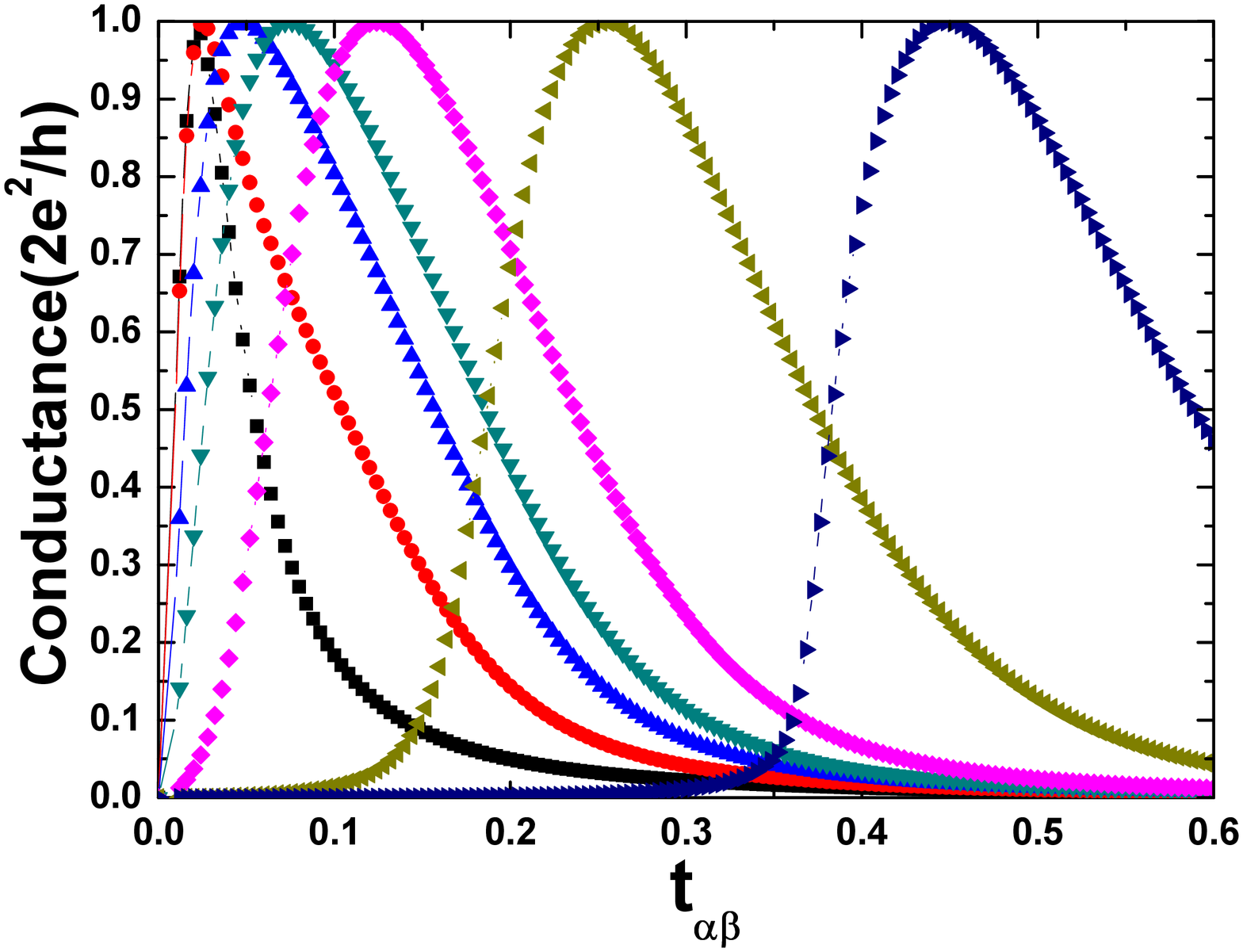}}}
\vskip0.0cm \caption{{(Color online) Conductance as a function of $t_{\alpha\beta}$ for $V_{g}/U=-0.5$ 
[(black) squares], $V_{g}/U=-0.14$ [(red) circles], $V_{g}/U=0.0$ [(blue) up-triangles], 
$V_{g}/U=0.08$ [(cyan) down-triangle], $V_{g}/U=0.22$ [(magenta) diamond], 
$V_{g}/U=0.6$ [left-triangle (gold)] and $V_{g}/U=1.2$ [(royal) right-triangle]. 
The other parameters are $t^\prime= 0.15$ and $U = 0.5$.}}  \label{condan}
\end{figure}

To investigate in more detail the effect of the different regimes in the transport properties  of
the DQD, in Fig.~\ref{condan} we present the conductance as a function of $t_{\alpha\beta}$ for
different values of $V_{g}$. These results are fitted by the analytic expression

\begin{eqnarray}
f(t_{\alpha\beta})=\frac{4{\Gamma}^{2}t_{\alpha\beta}^{2}}{({\Gamma}^{2}+t_{\alpha\beta}^{2}-
\tilde{\epsilon}_{i}^{2})^{2}-4\tilde{\epsilon}_{i}^{2}{\Gamma}^{2}},
\end{eqnarray}
where ${\Gamma}={\pi}t^{2}_{L(R)}{\rho}(E_{F})$ is the coupling constant and $i=\alpha (\beta)$. This
expression is an extension of the one obtained by Georges and Meir \cite{PhysRevLett.82.3508} and is valid also 
away from the particle-hole symmetric point, $V_{g}=-U/2$. Exactly at $V_{g}=-U/2$, the DQD has one
electron in each QD and the conductance is maximum when $t_{\alpha\beta}=\Gamma$.
For higher values of $t_{\alpha\beta}$, as $V_g$ moves away from $-U/2$, the DQD approaches the quarter filling 
molecular Kondo regime. 
This occurs when the charge of the system is nearly one ($n_{\alpha}+n_{\beta}{\approx}1$) or three electrons
($n_{\alpha}+n_{\beta}{\approx}3$). This regime in the conductance is reflected in Fig.~\ref{dp3}
by the two peaks which occur near the region of one and three electron occupation.
Figure~\ref{condan} shows the conductance as function of $t_{\alpha\beta}$ for
various values of $V_{g}$, that correspond to
the positions of the conductance peaks in Fig.~\ref{dp3} (for $V_{g}{\geq}-U/2$). From the results
obtained for $V_{g}>0$ one can identify two crossover regions for each $V_{g}$ as
$t_{\alpha\beta}$ increases: one corresponding to the transition from an empty dot situation to a
molecular Kondo regime, with an increasing conductance region and the other, for values of $t_{\alpha\beta}$ above the maximum
of the conductance, corresponding to the antiferromagnetic molecular regime where the conductance
decreases, and the DQD's occupancy increases.

To better understand the physics underlying the strong coupled QDs in the
transition region between the different regimes in the parameter space defined by $t_{\alpha\beta}$
and $V_{g}$,  we use the exact solution obtained in Appendix D, when the QDs are disconnected from
their leads. We obtain that $E_{0}^{(1e)}={\epsilon}_{0}-t_{\alpha\beta}$ and
$E^{(2e)}_0=2{\epsilon}_{0}+\frac{U-\sqrt{U^{2}+16t_{\alpha\beta}^{2}}}{2}$ are the lowest energies
in the one- and two-electron sector, respectively. Considering these 
expressions in a situation in which the charge occupation is increasing, we conclude that the inclusion
of an extra electron into the DQD requires an extra energy given by,
$E_{0}^{(2e)}-2E_{0}^{(1e)}=2t_{\alpha\beta}+\frac{U-\sqrt{U^{2}+16t_{\alpha\beta}^{2}}}{2}$.
This extra energy can be identified with an {\it effective Coulomb} interaction
\begin{eqnarray}
U_{eff}&=&2t_{\alpha\beta}+\frac{U-\sqrt{U^{2}+16t_{\alpha\beta}^{2}}}{2},
\end{eqnarray}
which, for small values of $U/t_{\alpha\beta}$, is $\approx U/2 + \mathcal{O}\left(U^2/16t_{\alpha\beta}\right)$.
The addition of a second electron in the DQD is achieved by adjusting the gate potential
$V_{g}$. In fact, for $V_{g} \approx t_{\alpha\beta} - U/2$, the DQD is basically double occupied, 
characterizing a transition from a molecular Kondo regime (at quarter-filling) to an
antiferromagnetic state, or to a two-impurity half-filling Kondo regime, depending upon the ratio $T_K/J$
being less or greater than unity, respectively.

\section{\label{app1} The minimization of the free energy in the SBMFA}

The free energy of the DQD is given by
\begin{eqnarray}
F({\gamma}_{l})&=&-k_BT\ln\left[\sum_{i}e^{-{\beta}E_{i}({\gamma}_{l})}\right],
\end{eqnarray}
where ${\beta}=1/k_BT$, and ${\gamma}_{l}$ denotes the
Lagrange multipliers. 
Differentiating F with
respect to ${\gamma}_{l}$ we obtain
\begin{eqnarray}
\frac{{\partial}F}{{\partial}{\gamma}_{l}}&=&\beta Tk_B\frac{{\sum_{i}}\frac{{\partial}E_{i}(
\gamma_l)}
{{\partial}(\gamma_l)}e^{-{\beta}E_{i}(\gamma_l)}}{\sum_{i}e^{-{\beta}E_{i}(\gamma_l)
}}\nonumber\\
&=&{\Big\langle}\frac{{\partial}E_{i}(\gamma_l)}{{\partial}{\gamma}_{l}}{\Big\rangle}\nonumber\\
& \approx &\frac{{\partial}{\langle}E_{i}(\gamma_l){\rangle}}{{\partial}{\gamma}_{l}},
\end{eqnarray}
where ${\gamma}_{l}$ denotes all the components of the $\gamma$-set 
defined in eq.~(\ref{888}). 
In addition, we have adopted an approximation assuming that the mean 
value of the derivative of the internal energy with respect to the 
components $\gamma_l$ is approximately equivalent to the derivative of the 
mean value of this energy with respect to these components.\cite{Cabrera:H-F}

\section{Conductance calculation}

The conductance is obtained from the Green's functions of the system. For the QD's
$\alpha(\beta)$ we have the local Green's functions
\begin{eqnarray}
G^{\sigma}_{{\alpha\alpha}(\beta\beta)}=\frac{\tilde{g}_{\alpha(\beta)\sigma}}{1-
t_{\alpha\beta}^{2}\tilde{g}_{\alpha\sigma}\tilde{g}_{\beta\sigma}},
\end{eqnarray}
obtained through a diagrammatic expansion which incorporates in the QD $\alpha(\beta)$ the physics
underlying the complete system, including the electron reservoirs. The function
$\tilde{g}_{\alpha(\beta)\sigma}$ that appears in this expression describes the sub-system composed by
the QD $\alpha(\beta)$ connected to the $L(R)$ reservoir. The calculation of this function results in 
\begin{eqnarray}
\tilde{g}_{\alpha(\beta)\sigma}&=&\frac{g_{\alpha(\beta)\sigma}}{1-t^{\prime2}g_{
\alpha(\beta)\sigma} \tilde{g}_{L(R)\sigma}},
\end{eqnarray}
where $g_{\alpha(\beta)\sigma}$ is the single-particle Green's function associated with the QD
$\alpha(\beta)$, while $\tilde{g}_{L(R)\sigma}=\frac{{\omega}-\sqrt{{\omega}^{2}-4t^{2}}}{2t^{2}}$ 
is the L(R) reservoir's Green's function projected onto its nearest QD $\alpha(\beta)$.

The non-local Green's functions of the system are given by 
\begin{eqnarray}
G^{\sigma}_{\alpha\beta(\beta\alpha)}&=&\frac{\tilde g_{\alpha(\beta)\sigma}t_{
\alpha\beta}\tilde g_{\beta(\alpha)\sigma}}
{1-t_{\alpha\beta}^{2}\tilde g_{\alpha(\beta)\sigma}\tilde g_{\beta(\alpha)\sigma}},
\end{eqnarray}
and
\begin{eqnarray}
G^{\sigma}_{{L\alpha}(R\beta)}&=&\frac{\tilde{g}_{L(R)\sigma}t^{\prime}\tilde{g}_{
\alpha(\beta)\sigma}}
{1-t_{\alpha\beta}^{2}\tilde{g}_{\alpha\sigma}\tilde{g}_{\beta\sigma}}
\end{eqnarray}
which are associated to charge transport between the QDs and from the $L(R)$ reservoir
to the QD $\alpha(\beta)$,  respectively. This propagators can be used to obtain the conductance $G$
of the  system. Using $G^{\sigma}_{\alpha\beta}$, for example, we obtain from the Keldysh formalism
\cite{Keldysh:1964ud} the following expression for the conductance:

\begin{eqnarray}
\label{conddddd}
G&=&\frac{2e^{2}}{{h\pi}^{2}}t_{L}^{2}t_{R}^{2}{\int}_{-\infty}^{+\infty}f_{L}(\omega)f_{R}(\omega){\mid}
G^{\sigma}_{\alpha\beta}(\omega){\mid}^{2}
\frac{{\partial}f_{L(R)}}{{\partial}\omega}d\omega,\nonumber\\
\end{eqnarray}
where $e$ is the electron charge, $h$ is Planck's constant, and $f_{L(R)}(\omega)$ the Fermi distribution
function associated to the L(R) reservoir.
\section{Exact Solution of isolated DQD at half-filling}

In order to better understand the characteristics of the antiferromagnetic and the Kondo
molecular  regime accessed when the DQD is at half- or quarter-filling, respectively, 
we calculate the exact solution when the DQD is decoupled from the metallic leads, as
shown in Fig.~(\ref{Apc1}). The exact solution obtained in this
Appendix allows the identification of the regions in the parameter space corresponding 
to different regimes and the crossovers between them.
\begin{figure}
\label{Apc1} \centering  \vskip0.0cm\hskip-0.0cm
\rotatebox{270}{\scalebox{0.3}{\includegraphics{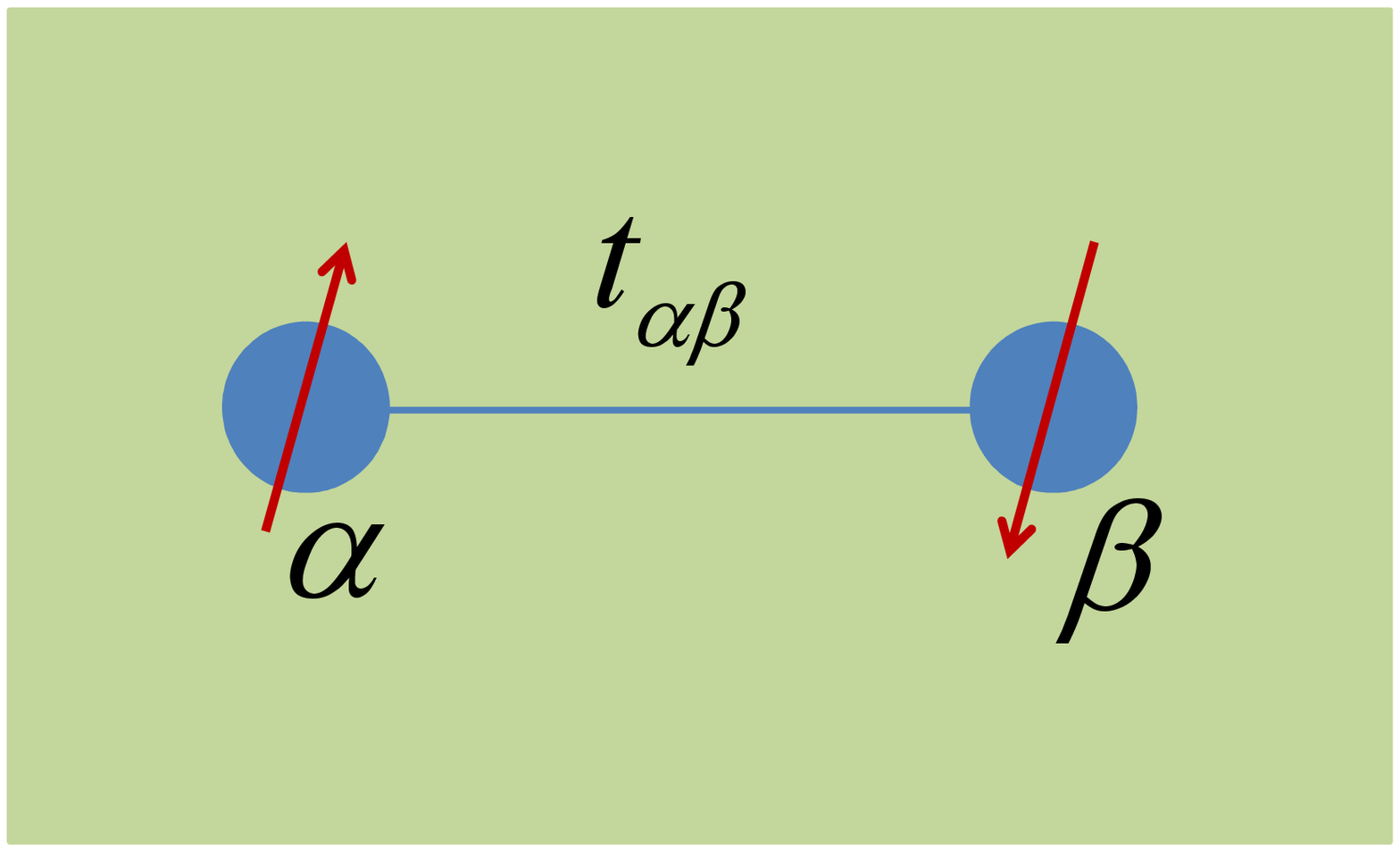}}}
\vskip-1.5cm \caption{(Color online) Schematic view of the decoupled DQD presented in Fig.~1.} \label{Apc1}
\end{figure}

The Hamiltonian for the decoupled DQD in Fig.~(\ref{Apc1}) is given by

\begin{eqnarray}
\label{aaaa}
H&=&\sum_{{i={\alpha},{\beta}}\atop{\sigma}}{\epsilon}_{i}n_{i\sigma}+\sum_{i={\alpha},{\beta}}
U_{i}n_{i\sigma}n_{i\bar{\sigma}}\nonumber\\
&&+\sum_{\sigma}t_{\alpha\beta}(c^{\dag}_{\alpha\sigma}c_{\beta\sigma}+c^{\dag}_{\beta\sigma}
c_{\alpha\sigma}),
\end{eqnarray}
where ${\epsilon}_{i}$ and $U_{i}$ are respectively the energy of the local state and the Coulomb
interaction in the i-th QD, $t_{\alpha\beta}$ is the hopping term and $\sigma$ the spin projections
of the electrons in the QDs. 
The DQD may have an electron occupancy of $N=1,2,3,4$. Thus, considering the system with
two electrons, $N=2$, we define the basis

\begin{eqnarray}
|{\varphi}_{1}{\rangle}&=&|{\uparrow}{\downarrow},0{\rangle}\\
|{\varphi}_{2}{\rangle}&=&|0,{\uparrow}{\downarrow}{\rangle}\\
|{\varphi}_{3}{\rangle}&=&\frac{1}{\sqrt{2}}[|{\uparrow},{\downarrow}{\rangle}-|{\downarrow},
{\uparrow}{\rangle}]\\
|{\varphi}_{4}{\rangle}&=&\frac{1}{\sqrt{2}}[|{\uparrow},{\downarrow}{\rangle}+|{\downarrow},
{\uparrow}{\rangle}]\\
|{\varphi}_{5}{\rangle}&=&|{\uparrow},{\uparrow}{\rangle}\\
|{\varphi}_{6}{\rangle}&=&|{\downarrow},{\downarrow}{\rangle},
\end{eqnarray}
where $|{\varphi}_{1}{\rangle}$, $|{\varphi}_{2}{\rangle}$ and $|{\varphi}_{3}{\rangle}$ are
states  with total spin $S_{T}=0$, while $|{\varphi}_{4}{\rangle}$, $|{\varphi}_{5}{\rangle}$ and
$|{\varphi}_{6}{\rangle}$ are states with $S_{T}=1$. Written in this basis, the Hamiltonian
can be separated into blocks, which correspond to the projections $S_{T}=0$ and $S_{T}=1$ of the total
spin. The block-matrix that corresponds to a spin projection $S_{T}=1$ is already diagonal and
its eigenvalues are $E_{4}=E_{5}=E_{6}=2{\epsilon}_{0}$ (for simplicity we consider
${\epsilon}_{\alpha}={\epsilon}_{\beta}={\epsilon}_{0}$). These energies are associated to 
states with $S_{T}=1$ and $S_{z}=1,0,-1$. 

The block-matrix corresponding to $S_{T}=0$ is given by 

$$ H^{AF}_{(2e)}=\left(
\begin{array}{cccc}
2{\epsilon}_{0}+U & 0 &-\sqrt{2}t_{\alpha\beta} \\
0 & 2{{\epsilon}_{0}}+U & \sqrt{2}t_{\alpha\beta} \\
-\sqrt{2}t_{\alpha\beta} & \sqrt{2}t_{\alpha\beta}& 2{\epsilon}_{0}\\
\end{array}
\right),$$
with the following eigenvalues: 
\begin{eqnarray}
{E}_{1}&=&2{\epsilon}_{0}+U\\
{E}_{2}&=&\frac{1}{2}[4{\epsilon}_{0}+U+\sqrt{U^{2}+16t_{\alpha\beta}^{2}}]\\
{E}_{3}&=&\frac{1}{2}[4{\epsilon}_{0}+U-\sqrt{U^{2}+16t_{\alpha\beta}^{2}}].
\end{eqnarray}
These results allow (as shown in the Appendix A) a better understanding of the processes 
underlying the transition between the different regimes of the DQD.

\bibliography{references}

\begin{thebibliography}{39}
\bibitem[{not({\natexlab{c}})}]{note2}
\bibinfo{note}{The results in Fig.~2(B) should be compared to those in Fig.~2
  of Ref.~\onlinecite{PhysRevB.62.10260}. Although the results presented there
  are for the even and odd `molecular orbitals', the density of states
  suppression has the same origin.}
\end{thebibliography}
\end{document}